\documentclass[
  reprint,
  aps,
  prl,
  amsmath,
  amssymb,
  superscriptaddress,
  nofootinbib,
  floatfix
]{revtex4-2}

\usepackage{mathtools}
\usepackage{bm}
\usepackage{graphicx}
\usepackage{xcolor}
\usepackage[hidelinks]{hyperref}

\newcommand{\dd}{\mathrm{d}}

\newcommand{\be}{\begin{equation}}
\newcommand{\e}{\end{equation}}
\newcommand{\beml}{\begin{subequations}}
\newcommand{\eml}{\end{subequations}}
\newcommand{\beq}{\begin{eqnarray}}
\newcommand{\eq}{\end{eqnarray}}
\newcommand{\ba}{\begin{array}}
\newcommand{\ea}{\end{array}}
\newcommand{\bpm}{\begin{pmatrix}}
\newcommand{\epm}{\end{pmatrix}}
\newcommand{\bc}{\begin{cases}}
\newcommand{\ec}{\end{cases}}
\newcommand{\lt}{\left}
\newcommand{\rt}{\right}
\newcommand{\n}{\nonumber}
\newcommand{\la}{\langle}
\newcommand{\ra}{\rangle}
\newcommand{\ep}{\varepsilon}
\newcommand{\bb}{\boldsymbol}
\newcommand{\bs}{\mathbf}

\newcommand{\0}{^{\phantom{\dagger}}}

\DeclareMathOperator{\Tr}{Tr}
\DeclareMathOperator{\En}{E}

\begin{document}

\title{Boundary Kerr Signatures of the Interband-Coherence Hall Effect}

\author{Ivan Iorsh}
\affiliation{Department of Physics, Engineering Physics \& Astronomy, Queen's University, Kingston, Canada}
\affiliation{Department of Physics, ITMO University, Saint-Petersburg, Russia}

\author{Mikhail Titov}
\affiliation{Institute for Molecules and Materials, Radboud University, Nijmegen, The Netherlands}

\date{\today}

\begin{abstract}
We identify a Hall response carried by optically induced interband coherence rather than by a non-equilibrium band population. In a weakly doped zinc-blende semiconductor, a longitudinal dc field drives a transverse flux of the conduction--valence coherence created by near-gap light. Angular averaging eliminates the homogeneous coherence density, while a lateral boundary converts the transverse flux into an antisymmetric, edge-localized, helicity-odd polarization. The resulting Kerr signal requires neither spin-orbit coupling nor a spin, valley, or orbital accumulation within an individual band. Within the eight-band Kane model, we derive the boundary kinetic equation and obtain a complex propagation length controlled by optical detuning and interband dephasing. The edge profile is monotonic at optical resonance and develops damped spatial oscillations away from it. The response is enhanced by electron--hole asymmetry and by strong interband mixing, making narrow-gap semiconductors especially favorable to observe the effect. These results establish dc-driven Kerr microscopy as a direct probe of an interband-coherence Hall effect and of multiband quantum kinetics in real space.
\end{abstract}

\maketitle

Scanning Kerr microscopy resolves current-induced changes of the helicity-odd optical susceptibility in real space. The seminal observation of opposite Kerr signals at the lateral edges of current-carrying GaAs channels was followed by optical studies of two-dimensional electron and hole gases, strained III--V semiconductors, and room-temperature structures~\cite{kato2004observation,wunderlich2005experimental,sih2005spatial,stern2006current}. These experiments established magneto-optical imaging as a powerful probe of electrically generated angular momentum and are commonly interpreted in terms of the spin Hall effect~\cite{dyakonov1971possibility,hirsch1999spin,murakami2003dissipationless,sinova2004universal,Engel2005Theory,Sinova2015review}. 

More recently, Kerr microscopy has detected current-induced orbital
accumulation in light transition metals%
~\cite{choi2023orbitalhall,lyalin2023chromium}, emphasizing that the
optical signal is not intrinsically tied to spin. These observations form
part of the broader development of orbitronics, in which electrically
generated orbital currents can propagate over long distances and can be
converted into spin angular momentum or magnetic torque at interfaces%
~\cite{go2021orbitronics,hayashi2023orbitaltransport}.

Conventional spin and orbital Hall effects are formulated as transverse currents of observables carried by Fermi-surface quasiparticles. Their edge accumulations are encoded in the dc correction to a diagonal density-matrix block. 

In particular, the orbital Hall effect can occur without spin-orbit coupling and is especially natural in multiorbital bands, including the six-dimensional valence-band sector of hole-doped semiconductors~\cite{bernevig2005orbitronics,go2018orbitaltexture,go2021orbitronics,cullen2026orbitalholes}. Its quantitative description in a crystalline multiorbital basis requires a consistent treatment of the position and velocity operators; recent Wannier-based calculations have emphasized the importance of the orbital-dependent anomalous
position for the intrinsic orbital Hall conductivity~\cite{go2024wannierorbitalhall}.

The mechanism proposed here is different: the optical field first creates an off-diagonal conduction--valence coherence, while the dc field then drives this coherence transversely. In this case, the measured Kerr response arises directly from the off-diagonal density matrix even if all dc-induced spin, valley, and orbital populations vanish.

%
%
\begin{figure}[!h]
  \centering
  \includegraphics[width=0.72\columnwidth]{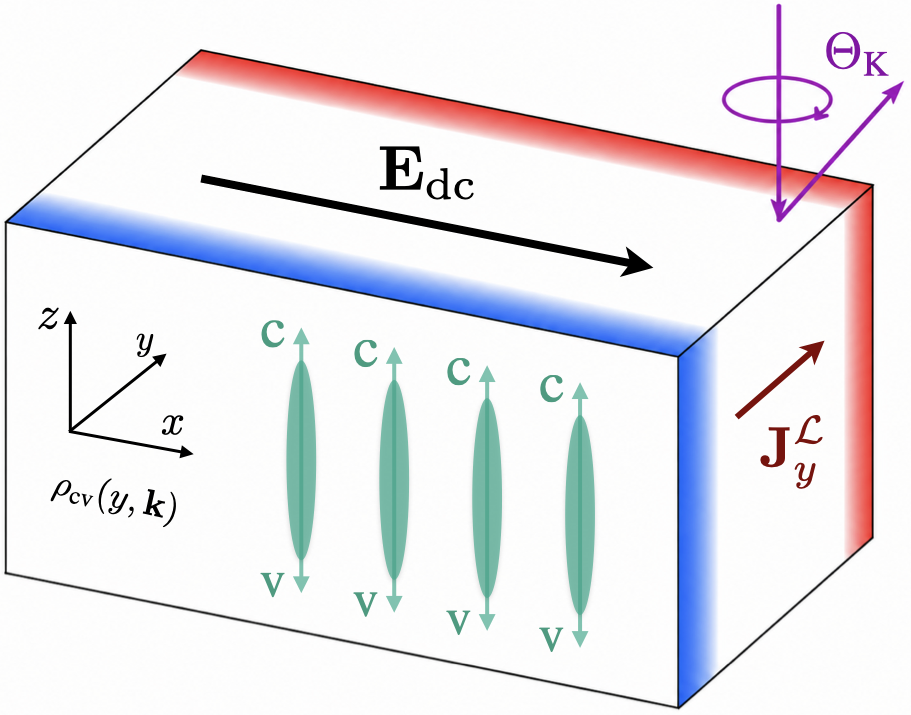}
  \caption{\textbf{Boundary conversion of interband-coherence flow into a Kerr signal.} Near-gap light propagating along $z$ creates the conduction--valence coherence $\rho_\textrm{cv}$. The dc field $E_{\rm dc}\hat{\bs x}$ drives its transverse flux $J_y^{\mathcal L}$ toward the side walls. The boundary normal $\pm\hat{\bs y}$ converts this flux into opposite helicity-odd edge polarizations, which rotate the reflected polarization by $\Theta_\textrm{K}$. The response is bilinear in the optical and dc fields and is not a spin or orbital population accumulation.}
  \label{fig:setup}
\end{figure}

Interband coherence is a central component of electric-field response in multiband crystals. In quantum kinetic theory it accompanies band repopulation and generates intrinsic Berry-phase terms, together with disorder-mediated corrections, in transverse transport~\cite{nagaosa2010anomalous,ado2015,ado2016,culcer2017interband,culcer2026}. Off-diagonal density-matrix elements are likewise fundamental to the nonlinear optical response and semiconductor Bloch equations~\cite{lindberg1988bloch,aversa1995lengthgauge,sipe2000second}. The effect considered here combines these two settings: near-gap light creates a conduction--valence polarization, and a longitudinal dc field produces its momentum-space drift. The component odd in transverse momentum carries a finite transverse flux but no homogeneous helicity; a lateral boundary converts that flux into a local helicity-odd optical polarization.

We call this mixed dc--optical response the \emph{interband-coherence Hall effect}. Figure~\ref{fig:setup} summarizes the mechanism. A field $E_{\rm dc}\hat{\bs x}$ drives a coherence flux toward the side walls, while light propagating along $\hat{\bs z}$ creates and probes the interband polarization. The boundary normal $\pm\hat{\bs y}$ fixes the sign of the local response, so opposite edges carry opposite Kerr signals. Here ``interband coherence'' refers specifically to the off-diagonal block $\rho_\textrm{cv}$ in the Bloch basis that diagonalizes the field-free Hamiltonian.

We formulate the effect for a weakly doped zinc-blende semiconductor within the eight-band Kane model~\cite{kane1957band,voon2009kp,yu2010fundamentals}. The boundary problem produces a complex coherence length governed by optical detuning and the interband dephasing rate. At resonance, the Kerr profile decays monotonically from a side wall; away from resonance, it develops damped spatial oscillations. Below we focus on the intrinsic Berry-connection contribution. A controlled treatment of side-jump and skew-scattering terms requires a specified disorder correlator and is not universal, as is familiar from the anomalous Hall effect~\cite{nagaosa2010anomalous,ado2015,ado2016,ado2017correlations,culcer2026}.

\paragraph{Optical response and kinetic equation.}
For a linearly polarized probe, the Kerr rotation $\Theta_\textrm{K}$ and ellipticity $\eta_\textrm{K}$ are determined by the difference $\delta\chi=\chi_{++}-\chi_{--}$ between the susceptibilities for right- and left-circular polarization. In the weak-absorption limit, one finds
\be
  \Theta_\textrm{K}=-2\pi\frac{\operatorname{Im}\delta\chi}{\sqrt{\ep}(\ep-1)},
  \qquad
  \eta_\textrm{K}=2\pi\frac{\operatorname{Re}\delta\chi}{\sqrt{\ep}(\ep-1)},
  \label{eq:kerr}
\e
where $\varepsilon=1+2\pi(\chi_{++}+\chi_{--})$ is the background dielectric function. Equation~\eqref{eq:kerr} will be used to convert the calculated circular susceptibility into the observable rotation and ellipticity.

In the field-free Bloch eigenbasis $|n\bs{k}\rangle$, the helicity-odd susceptibility $\delta\chi=\delta\chi(y,\omega)$ is
\be
  \delta\chi=\frac{ie}{\omega\mathcal E}
  \sum_{\xi=\pm}\xi\!\int\!\frac{\dd^3 \textrm{k}}{(2\pi)^3}
  \Tr\!\left[\rho_{\textrm{cv},\xi}(y,\bs{k};\omega)\,
  \mathcal V_{\textrm{vc},\xi}(\bs{k})\right],
  \label{eq:susceptibility}
\e
where $\mathcal E$ is the optical-field amplitude, $\xi=\pm$ labels its helicity, and $\mathcal V_{\textrm{vc},\xi}$ is the corresponding circular component of the interband velocity matrix. We set $\hbar=1$ throughout the main text and End Matter.

The density matrix obeys the semiconductor Bloch equation in the length gauge,
\begin{equation}
  \partial_t\rho+i[H_0-e\,\bs{E}\!\cdot\!\bs{r},\rho]
  +\frac{1}{2}\{\mathcal V_y,\partial_y\rho\}=\mathcal I[\rho],
  \label{eq:sbe}
\end{equation}
where $H_0$ is the field-free Hamiltonian and $\mathcal I$ is the collision integral. The anticommutator with the velocity operator $\mathcal V_y =i [\hat{r}_y,H_0]$ is included since the density matrix acquires an appreciable dependence on the transversal coordinate $y$. In the setup of Fig.~\ref{fig:setup} the translational invariance is broken only along $y$, so Eq.~\eqref{eq:sbe} must be supplemented by boundary conditions at $y=0$ and $y=W$.

The position operator $\hat{\bs{r}}$ is represented by its commutator with an arbitrary single-particle operator $\mathcal O$,
\beml
\label{eq:covariant_derivative}
\begin{align}
  \langle n\bs{k}|[\hat{r}_j,\mathcal O]|m\bs{k}'\rangle
  &=\delta(\bs{k}-\bs{k}')\,(\mathcal D_j\mathcal O)_{nm},\\
  \mathcal D_j\mathcal O
  &=i\partial_{k_j}\mathcal O+[\mathcal A_j,\mathcal O],
\end{align}
\eml
where $\mathcal A_j=i\mathcal R^\dagger\partial_{k_j}\mathcal R$ is the Berry connection and $\mathcal R$ diagonalizes $H_0$. We denote its conduction and valence blocks by $\mathcal A_\textrm{cc}$, $\mathcal A_\textrm{cv}$, $\mathcal A_\textrm{vc}$, and $\mathcal A_\textrm{vv}$.

The total field is a superposition $\bs{E}=\mathcal E\hat{\bs e}_\xi e^{-i\omega t}+E_{\rm dc}\hat{\bs x}$, hence we can expand the density matrix as
\be
\rho\!=\!\rho^{(0,0)}\!+\!eE_{\rm dc}\rho^{(0,1)}\!+\!e\mathcal E\rho^{(1,0)}\!e^{-i\omega t}\!+\!e^2E_{\rm dc}\mathcal E\rho^{(1,1)}\!e^{-i\omega t},
  \label{eq:expansion}
\e
where the first superscript counts powers of the optical field and the second counts powers of the dc field. The equilibrium term is $\rho^{(0,0)}_{nm}=f(\epsilon_n)\delta_{nm}$, with $f$ the Fermi--Dirac distribution. The mixed term $\rho^{(1,1)}$ determines the current-induced Kerr signal.

Two physically distinct sources contribute to $\rho^{(1,1)}$. The first is the optical readout of the dc-induced diagonal distribution, $\rho^{(0,1)}_\textrm{cc}$ or $\rho^{(0,1)}_\textrm{vv}$, including conventional spin and orbital Hall accumulations. The second is the dc drift of the optical coherence $\rho_\textrm{cv}^{(1,0)}$. This second source may remain finite without spin-orbit coupling and without a dc correction to any diagonal-band angular momentum.

\paragraph{Spin-density channel.}
For comparison, consider a weakly $n$-doped semiconductor with a prescribed conduction-band spin density $s_z(y)$. Keeping the resonant interband term, the corresponding mixed optical coherence is
\be
  \rho_{\textrm{cv},\xi}^{(1,1)}(\bs{k})
  =-\rho_\textrm{cc}^{(0,1)}(\bs{k})\,\mathcal A_{\textrm{cv},\xi}(\bs{k})\,G_0(\omega,\bs{k}),
  \label{eq:spin_coherence}
\e
where $G_0^{-1}(\omega,\bs{k})=\Delta_{\bs{k}}-\omega+i\gamma_\phi$, $\Delta_{\bs{k}}=\epsilon_\textrm{c}(\bs{k})-\epsilon_\textrm{v}(\bs{k})$ is the interband transition energy, and $\gamma_\phi=\tau_\phi^{-1}$ is the homogeneous interband dephasing rate. In the spherical Kane approximation at low temperature, Eq.~\eqref{eq:susceptibility} gives
\be
  \delta\chi^{\rm spin}(y,\omega)
  =\frac{2e^2\mathcal P^2}{3\omega E_\textrm{g}}
  \frac{s_z(y)}{\delta_\textrm{F}+i\gamma_\phi},
  \label{eq:spin_channel}
\e
where $\mathcal P$ is the Kane matrix element, $E_\textrm{g}$ is the band gap, and $\delta_\textrm{F}=\Delta_{k_\textrm{F}}-\omega$ is the detuning at the Fermi wave vector $k_\textrm{F}$. The normalization of $s_z$ includes the density-of-states and thermal-shell factors, so Eq.~\eqref{eq:spin_channel} specifies only the optical conversion of an already established spin accumulation; its microscopic generation lies outside the scope of the present analysis.

\paragraph{Interband-coherence drift.}
The optical field creates
\be
  \rho_{\textrm{cv},\xi}^{(1,0)}=(1-f_\textrm{c})\mathcal A_{\textrm{cv},\xi}G_0,
  \label{eq:optical_coherence}
\e
where $f_\textrm{c}=f[\epsilon_\textrm{c}(\bs{k})]$. Retaining the intrinsic dc drift of Eq.~\eqref{eq:optical_coherence}, the $\textrm{cv}$ block of Eq.~\eqref{eq:sbe} becomes
\begin{align}
  &-iG_0^{-1}\rho_{\textrm{cv},\xi}^{(1,1)}
  +\frac{1}{2}\left[
  \mathcal V_{\textrm{cc},y}\partial_y\rho_{\textrm{cv},\xi}^{(1,1)}
  +(\partial_y\rho_{\textrm{cv},\xi}^{(1,1)})\mathcal V_{\textrm{vv},y}
  \right]\nonumber\\
  &\hspace{39mm}=i\lt(\mathcal D_x\rho_\xi^{(1,0)}\rt)_\textrm{cv},
  \label{eq:coherence_kinetic}
\end{align}
which generates the combination $\gamma_\phi-i\delta_\textrm{F}$ governing spatial decay and phase accumulation.

To obtain a transparent analytic result, we first set the valence-band spin-orbit splitting $\Delta_{\rm so}$ to zero. Then $G_0$ is scalar in the cc and vv subspaces and the velocity blocks are trivial, $\mathcal V_{\textrm{cc},y}=(k_y/m_\textrm{c})\openone_\textrm{c}$ and $\mathcal V_{\textrm{vv},y}=-(k_y/m_\textrm{v})\openone_\textrm{v}$, where $m_\textrm{c}$ and $m_\textrm{v}$ are positive conduction-electron and valence-hole masses. To leading order in $E_\textrm{g}^{-1}$, the relevant Berry connections are $\mathcal A_\textrm{vv}=[\mathcal P^2/(2E_\textrm{g}^2)](\bs{k}\times\bs{L})\otimes\openone_s$ and $\mathcal A_\textrm{cc}=0$, where $\bs{L}$ is the orbital momentum operator. We project Eq.~\eqref{eq:coherence_kinetic} onto the helicity-odd optical channel by defining
\be
  \mathcal L_{\bs{k}}(y)=
  \sum_{\xi=\pm}\xi\,
  \Tr\!\lt[\rho_{\textrm{cv},\xi}^{(1,1)}(y,\bs{k})
  E_\textrm{g}\mathcal A_{\textrm{vc},\xi}(\bs{k})\rt],
  \label{eq:L_definition}
\e
where $\mathcal A_{\textrm{vc},\xi}=\mathcal A_{\textrm{cv},\xi}^\dagger$. Equation~\eqref{eq:coherence_kinetic} then reduces to
\be
  G_0^{-1}\mathcal L_{\bs{k}}
  +i(k_y/\mu)\partial_y\mathcal L_{\bs{k}}
  =2(\mathcal P^4/E_\textrm{g}^3)k_y(1-f_\textrm{c})G_0,
  \label{eq:L_kinetic}
\e
where $\mu^{-1}=(m_\textrm{c}^{-1}-m_\textrm{v}^{-1})/2$. Its right hand side is odd in $k_y$ and therefore drives a transverse flux without producing a homogeneous helicity density. In the equal-mass limit, $m_\textrm{c}=m_\textrm{v}$, the relative drift velocity $u_y=k_y/\mu$ vanishes and the boundary layer collapses. Although the formal pointwise value at an ideal mathematical interface is nonuniform in this limit, any integrated or finite-resolution Kerr signal vanishes. 

We interpret the isotropic moment $\mathcal L(y)=\lt\la\mathcal L_{\bs{k}}\rt\ra$ as an interband coherence density and its transverse flux $J_y^{\mathcal L}(y)=\lt\la u_y\mathcal L_{\bs{k}}\rt\ra$ as the interband coherence flux (the Hall coherence current), where $u_y=k_y/\mu$. The angular brackets denote here an average over the angles of $\bs{k}$ on the resonant shell. 

Both $\mathcal L$ and $J_y^{\mathcal L}$ are complex amplitudes at the optical frequency. Their real and imaginary parts represent optical observables rather than usual density and current. The two-moment closure of Eq.~\eqref{eq:L_kinetic} results, nevertheless, in the drift-diffusion equations on the coherence flux and density, 
\beml
\label{eq:moment_equations}
\begin{align}
  (\gamma_\phi-i\delta_\textrm{F})\mathcal L+\partial_yJ_y^{\mathcal L}&=0,\\
  (\gamma_\phi-i\delta_\textrm{F})J_y^{\mathcal L}
  +\frac{u_\textrm{F}^2}{3}\partial_y\mathcal L
  &=-\frac{2i\mathcal P^4k_\textrm{F}^2}{3\mu E_\textrm{g}^3}(1-f_\textrm{c})G_0,
  \label{eq:Diffusion}
\end{align}
\eml
where $u_\textrm{F}=k_\textrm{F}/\mu$ is the relative conduction--valence drift velocity on the resonant Fermi shell rather than the conduction-band Fermi velocity.  In order to derive Eqs.~(\ref{eq:moment_equations}) we had to use  $\lt\la k_y^2\mathcal L_{\bs{k}}\rt\ra = \lt\la k_y^2\rt\ra \lt\la \mathcal L_{\bs{k}}\rt\ra$, that gives a qualitative diffusion-like description but is violated in practice in the boundary layer, where the density matrix $y$ dependence is too strong.

%
%
\begin{figure}[t]
  \centering
  \includegraphics[width=0.98\columnwidth]{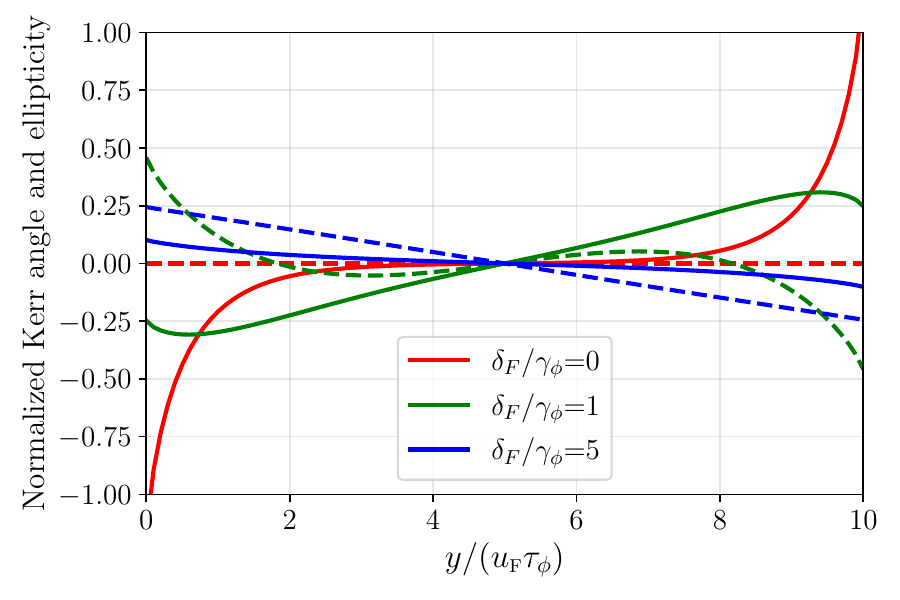}
  \caption{\textbf{Coherent edge propagation revealed by optical detuning.} Normalized Kerr angle (solid lines) and ellipticity (dashed lines) profiles are shown versus distance. The resonant signal decays monotonically, whereas detuning gives the propagation length in Eq.~\eqref{eq:coherence_length} an imaginary part and produces damped sign changes. Such oscillations are absent from a purely relaxational spin-diffusion profile. The slab thickness $W=10\, u_\textrm{F}\tau_\phi$.}
  \label{fig:spatial}
\end{figure}

Closed boundary conditions, $\lt.J_{y}^{\mathcal L}\rt|_{y=0}=\lt.J_{y}^{\mathcal L}\rt|_{y=W}=0$ make the solution of Eqs.~\eqref{eq:moment_equations} be controlled by the complex length scale $u_\textrm{F}/(\gamma_\phi-i\delta_\textrm{F})$. At resonance, this length is real and quantifies an exponential decay, while finite detuning generates damped oscillations.
spatial
Let us now turn to a more microscopic treatment of Eq.~(\ref{eq:L_kinetic}) without postulating Eqs.~(\ref{eq:moment_equations}). A quantitative spatial profile follows from the momentum-resolved Eq.~\eqref{eq:L_kinetic}. We impose diffuse Fuchs--Sondheimer conditions at $y=0,W$~\cite{fuchs1938conductivity,sondheimer2001mean,Okulov1979_SovJLowTempPhys}. With $\mathcal L_{\pm,\bs{k}}=\mathcal L_{\bs{k}}|_{k_y\gtrless0}$, the outgoing distribution equals the flux-weighted angular average of the incoming distribution,
\beml
\label{eq:fuchs_bc}
\begin{align}
&\mathcal{L}_{+,\mathbf{k}}|_{y=0} = \frac{\int_{k_y<0}  d\Omega_{\bf k}|u_y|\mathcal{L}_{-,\mathbf{k}}|_{y=0}}{\int_{k_y<0} d\Omega_{\bf k} |u_y|},\\ & \mathcal{L}_{-,\mathbf{k}}|_{y=W} = \frac{\int_{k_y>0} d\Omega_{\bf k}|u_y|\mathcal{L}_{+,\mathbf{k}}|_{y=W}}{\int_{k_y>0}d\Omega_{\bf k} |u_y|},
\end{align}
\eml
where $\mathcal{L}_{+ {\bf k}}$ corresponds to the flux away from the boundary $y=0$ and towards $y=W$ and $\mathcal{L}_{-, {\bf k}}$ corresponds to the flux in the opposite direction. The conditions of Eq.~(\ref{eq:fuchs_bc}) also imply  $\lt.J_{y}^{\mathcal L}\rt|_{y=0}=\lt.J_{y}^{\mathcal L}\rt|_{y=W}=0$. Equations (\ref{eq:fuchs_bc}) model fully diffuse but coherence-preserving reflection. Additional surface dephasing can be incorporated through a reduced outgoing coherence amplitude.

For a wide slab, $W\gg u_\textrm{F}/\gamma_\phi$, the intrinsic interband-coherence contribution to the circular susceptibility reads
\begin{align}
  \delta\chi^{\rm IC}(y,\omega)={}&
  \frac{2e^2\nu_\textrm{F}\mathcal P^2}{3\omega E_\textrm{g}}
  \frac{3 eE_{\rm dc}\lambda_\textrm{F}}{2\pi(\delta_\textrm{F}+i\gamma_\phi)}
  \left(\frac{\mathcal Pk_\textrm{F}}{E_\textrm{g}}\right)^2\nonumber\\
  &\times
  \left[F\left((W-y)/\ell_{\phi} \right)-F\left(y/\ell_{\phi}\right)\right],
  \label{eq:IC_susceptibility}
\end{align}
where we defined the complex interband dephasing length
\be
  \ell_\phi=u_\textrm{F}/(\gamma_\phi-i\delta_\textrm{F}).
  \label{eq:coherence_length}
\e
Here, $\nu_\textrm{F}$ is the joint density of states on the resonant shell, $\lambda_\textrm{F}=2\pi/k_\textrm{F}$,  $F(x)=\En_3(x)+(2/3)\En_2(x)$, and
\be
  \En_n(x)=\int_1^\infty \dd t\,e^{-xt}t^{-n}
  \label{eq:exp_integral}
\e
is the generalized exponential integral. The importance of a boundary-controlled dephasing
length for a dc-driven Kerr response was anticipated in Ref.~\cite{voss2025orbitalmagnetization}.

The characteristic length emerging from Eqs.~\eqref{eq:Diffusion} is smaller than Eq.~\eqref{eq:coherence_length} by a factor of $\sqrt{3}$. This discrepancy can be traced back to a breakdown of the approximation $\langle k_y^2 \mathcal{L}_{\bf k} \rangle\approx \langle k_y^2\rangle \langle \mathcal{L}_{\bf k} \rangle$ used in Eqs.~\eqref{eq:Diffusion}. 

For a specular wall, $\mathcal L_{+,\bs{k}}=\mathcal L_{-,\bs{k}}$ at the same boundary, Eq.~\eqref{eq:IC_susceptibility} retains its form but $F(x)$ is replaced by $2\En_3(x)$. The edge value changes from $F(0)=7/6$ for the diffuse wall to $2\En_3(0)=1$ for the specular wall, while the integrated accumulation  $\int dy\, F(y/\ell_{\phi})$ is identical in both cases. Thus, boundary roughness modifies the local Kerr amplitude without changing the total boundary conversion. Equation~\eqref{eq:IC_susceptibility} is odd under $y\mapsto W-y$ and reverses sign with $E_{\rm dc}$, as required for a boundary Hall response.

The coherence signal differs qualitatively from the spin-density channel in Eq.~\eqref{eq:spin_channel}. Its characteristic length is set by interband dephasing length $u_\textrm{F}\tau_\phi$ rather than spin diffusion, and the complex length in Eq.~\eqref{eq:coherence_length} produces spatial oscillations when $|\delta_\textrm{F}|\gtrsim\gamma_\phi$. Figure~\ref{fig:spatial} shows that the profile changes from monotonic decay at resonance to a sequence of damped sign reversals away from resonance, providing a direct real-space fingerprint of coherent propagation.

\begin{figure}[t]
  \centering
  \includegraphics[width=0.98\columnwidth]{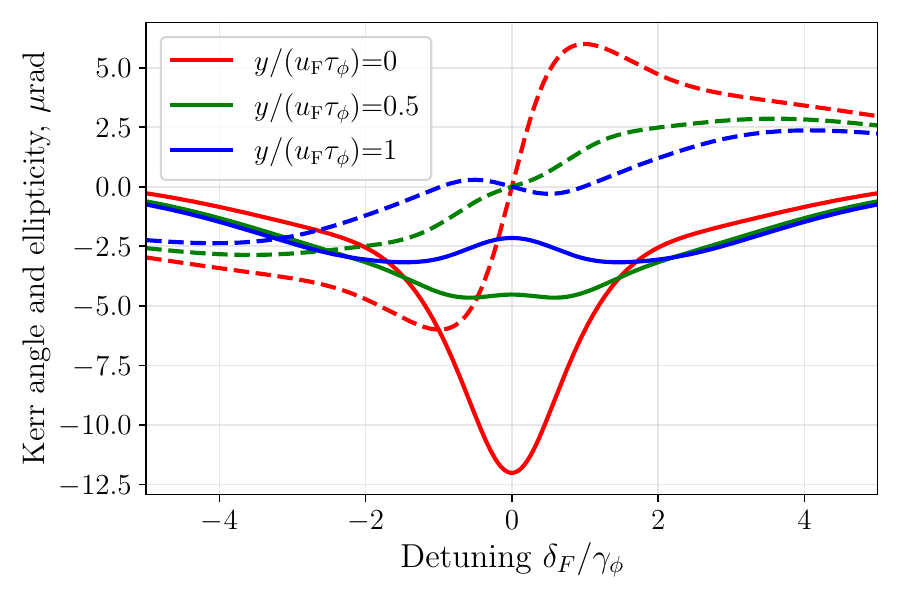}
  \caption{\textbf{Resonantly enhanced Kerr (solid lines) and ellipticity (dashed lines) response in narrow-gap InSb.} Kerr rotation and ellipticity in the vicinity of the side wall are plotted versus detuning for $k_\textrm{F}=10^8\,\mathrm{m}^{-1}$, $E_{\rm dc}=10\,\mathrm{mV}\,\mu\mathrm{m}^{-1}$, $T<10\,\mathrm{K}$, and the illustrative for the dephasing time $\tau_\phi=\gamma_\phi^{-1}=1\,\mathrm{ps}$, which is a very conservative estimate. The line shapes follow from the complex susceptibility: rotation is predominantly absorptive near resonance, while ellipticity changes sign across it.}
  \label{fig:insb}
\end{figure}

The amplitude is governed by the interband-mixing parameter $\mathcal Pk_\textrm{F}/E_\textrm{g}$ and does not require spin-orbit coupling. This distinguishes the effect from a spin Hall accumulation and from an orbital Hall accumulation encoded in a diagonal valence-band density matrix. The inverse powers of $E_\textrm{g}$ in Eq.~\eqref{eq:IC_susceptibility} favor narrow-gap semiconductors. For representative InSb parameters, $\mathcal Pk_\textrm{F}/E_\textrm{g}$ can approach unity, although quantitative work in that regime should eventually replace the leading $E_\textrm{g}^{-1}$ expansion by a numerical treatment of the full Kane Hamiltonian. Figure~\ref{fig:insb} illustrates the predicted dispersive Kerr angle and ellipticity near the absorption edge.

%
%

Short-range disorder adds side-jump and skew-scattering corrections to Eq.~\eqref{eq:coherence_kinetic}. Their relative weights depend on the disorder correlator, and rare closely spaced impurities can contribute at the same order as noncrossing terms~\cite{nagaosa2010anomalous,ado2015,ado2016,ado2017correlations,culcer2026}. We therefore cannot formulate a universal scattering correction to Eq.~\eqref{eq:IC_susceptibility}. Smooth correlated disorder should suppress the most singular short-range contributions. Another strategy is to replace the dc field by a sub-gap ac drive whose frequency exceeds inverse transport time. In analogy with finite-frequency anomalous Hall measurements, this may isolate the intrinsic response~\cite{kim2010infraredAHE}.

Polarization-resolved near-field magneto-optical microscopy offers an excellent underexplored tool for observing the effect~\cite{fumagalli1998quantitative}. Even without resolving the full profile, current reversal, opposite signs at opposite edges, the detuning-dependent phase, and the sign-changing spatial structure provide a combined experimental test of the proposed mechanism.

\paragraph{Symmetry and electron--hole asymmetry.}
The interband coherence response does not require a spin-orbit-generated axial vector. Its local sign is fixed by three mutually perpendicular polar vectors: the dc-driving direction $\hat{\bs{x}}$, the optical propagation direction $\hat{\bs{z}}$, and the boundary normal ${\bs{n}}=\pm\hat{\bs{y}}$. The pseudoscalar ${\bs{n}}\!\cdot\!(\hat{\bs{z}}\times{\bs{E}}_{\rm dc})$ changes sign under current reversal and between opposite edges, exactly as in Eq.~\eqref{eq:IC_susceptibility}. A second requirement is conduction--valence asymmetry. In the parabolic model it enters through $m_\textrm{c}\neq m_\textrm{v}$, which gives a nonzero relative coherence velocity $u_\textrm{F}$. Thus, the driven current operationally breaks the time-reversal symmetry, while the boundary selects one of the two opposite normals and thereby removes the local inversion symmetry. The Hall-like handedness is supplied by the driven boundary geometry rather than by magnetic order.

\paragraph{Relation to the orbital Hall effect.}
The distinction to the orbital Hall effect is most transparent in a $p$-doped semiconductor. When the Fermi energy lies in the valence band, the six-dimensional $\Gamma_{8v}\oplus\Gamma_{7v}$ sector supports quasiparticles with a nontrivial orbital matrix structure. A conventional orbital Hall effect then produces a transverse orbital current and an edge orbital density encoded in the diagonal dc correction $\rho_\textrm{vv}^{(0,1)}$~\cite{bernevig2005orbitronics,go2018orbitaltexture,cullen2026orbitalholes}. Kerr microscopy can probe this orbital density, as demonstrated in metallic orbitronics~\cite{choi2023orbitalhall,lyalin2023chromium}. The interband-coherence contribution instead originates from the drift of $\rho_\textrm{cv}^{(1,0)}$ and enters $\rho_\textrm{cv}^{(1,1)}$. It therefore survives as a separate channel in a hole-doped system, although its matrix structure will be richer and its propagation length will remain controlled by interband dephasing rather than by orbital diffusion. One may describe the resulting matrix-resolved response as an orbital-resolved interband-coherence Hall effect, which may extend to tens of microns from the boundary. However, this effect should not be confused with the orbital Hall accumulation.

\paragraph{Interband dephasing scale.}
The parameter $\gamma_\phi$ is the homogeneous linewidth of the optical conduction--valence polarization, where $\tau_\phi=\gamma_\phi^{-1}$ is its phase-memory time. This is not the momentum-transport time of the conduction electrons. In semiconductor Bloch equations the polarization dephasing contains population decay and pure-dephasing contributions~\cite{lindberg1988bloch}. Carrier--carrier collisions, phonons, radiative decay, and disorder that acts differently on the conduction and valence states can all contribute to $\tau_\phi$. Four-wave-mixing experiments show that exciton--exciton and carrier scattering can dominate at finite excitation density, acoustic and optical phonons control the temperature dependence in clean structures, and localized defects can strongly shorten the interband coherence in low-temperature-grown GaAs~\cite{schultheis1986dephasing,borri1999phonon,webber2018dephasing}. In contrast, coherent interband dynamics in bulk InSb has been observed directly with field-resolved four-wave mixing~\cite{junginger2012insb}.

For the weak-probe, low-density, low-temperature regime relevant here, $\tau_\phi$ may therefore exceed the illustrative $1\,\mathrm{ps}$ used in Fig.~\ref{fig:insb}, particularly in a clean sample where inelastic phonon processes rather than impurity momentum relaxation set the homogeneous linewidth. Smooth static disorder can broaden the optical transition inhomogeneously without producing the same local phase decay, whereas band-dependent short-range disorder contributes to genuine homogeneous dephasing. A longer $\tau_\phi$ increases both the resonant edge amplitude and the propagation distance $u_\textrm{F}\tau_\phi$. Quantitative predictions should therefore use the near-gap optical linewidth or a four-wave mixing measurement on the same sample, rather than identifying $\tau_\phi$ with the transport lifetime.

\paragraph{Conclusion.}
We have proposed an intrinsic boundary Kerr response generated by the dc drift of optically induced conduction--valence coherence. The response is transverse, current odd, and antisymmetric between opposite edges, yet it requires neither spin-orbit coupling nor a diagonal-band spin, valley, or orbital accumulation. Its existence follows from the handed boundary geometry and from electron--hole asymmetry, while its spatial structure is governed by the interband phase-memory time. The theory predicts a monotonic resonant profile, detuning-induced damped oscillations, sensitivity of the local amplitude to boundary scattering, and strong enhancement in narrow-gap multiband semiconductors. It also establishes a sharp distinction between the conventional orbital Hall effect, which acts through a diagonal valence-band distribution, and the interband-coherence Hall effect, which probes the off-diagonal optical density matrix. These signatures make dc-driven Kerr microscopy a direct real-space probe of interband quantum kinetics and open the route to measuring coherence transport that is invisible to conventional charge, spin, or orbital accumulation measurements.

\begin{acknowledgments}
The authors thank Marcos Guimaraes and Yurii Mokrousov for useful discussions.
\end{acknowledgments}

\bibliographystyle{apsrev4-2}
\bibliography{Boundary_Kerr_Interband_Coherence_v6}

\clearpage
\section{END MATTER}

\paragraph{Berry connections in the spherical Kane model.}
For completeness, we collect the Berry connections used in the main text. The calculation retains the leading non-vanishing terms in $E_\textrm{g}^{-1}$ and assumes equal light- and heavy-hole masses in the spherical approximation. The eight-band Kane Hamiltonian is first block-diagonalized by a Schrieffer--Wolff transformation and is then diagonalized within the $\Gamma_{6c}$ conduction subspace and the $\Gamma_{8v}\oplus\Gamma_{7v}$ valence subspace. The following matrices are written on the same field-free eigenbasis as Eq.~\eqref{eq:covariant_derivative}.

The conduction-band Berry connection is
\begin{equation}
  \mathcal A_{\textrm{cc},i}=-\frac{\mathcal P^2}{3E_\textrm{g}^2}
  \left[1-\frac{1}{(1+\Delta_{\rm so}/E_\textrm{g})^2}\right]
  (\bs{k}\times{\bb\sigma})_i,
  \label{eq:Acc_end}
\end{equation}
where $\bb{\sigma}$ is the vector of Pauli matrices that acts on the $\Gamma_{6c}$ Kramers doublet. This connection vanishes when the valence-band spin-orbit splitting $\Delta_{\rm so}$ is set to zero.

The valence-band connection is
\begin{equation}
  \mathcal A_{\textrm{vv},i}=-\frac{\mathcal P^2}{2E_\textrm{g}^2}\epsilon_{ijl}k_j
  \begin{pmatrix}
    \frac{2}{3}J_l^{(3/2)} & \frac{T_l}{1+\Delta_{\rm so}/E_\textrm{g}}\\[2pt]
    \frac{T_l^\dagger}{1+\Delta_{\rm so}/E_\textrm{g}} &
    \frac{4\sigma_l}{3(1+\Delta_{\rm so}/E_\textrm{g})^2}
  \end{pmatrix}
  \label{eq:Avv_end}
\end{equation}
where $J_l^{(3/2)}$ is the spin-$3/2$ angular-momentum matrix in the $\Gamma_{8v}$ sector, while $\sigma_l$ acts in the $\Gamma_{7v}$ split-off doublet. The $4\times2$ matrices connecting these sectors are
\begin{equation}
  \resizebox{0.98\columnwidth}{!}{$
  \displaystyle
  \begin{aligned}
  T_x&=\frac{1}{\sqrt6}
  \begin{pmatrix}
    1&0\\0&-1/\sqrt3\\-1/\sqrt3&0\\0&1
  \end{pmatrix},
  &
  T_y&=-\frac{i}{\sqrt6}
  \begin{pmatrix}
    1&0\\0&-1/\sqrt3\\1/\sqrt3&0\\0&-1
  \end{pmatrix},\\
  T_z&=\frac{\sqrt2}{3}
  \begin{pmatrix}
    0&0\\-1&0\\0&1\\0&0
  \end{pmatrix}.
  \end{aligned}
    \label{eq:T_end}
  $}
\end{equation}
These conventions fix the ordering of rows and columns in Eq.~\eqref{eq:Avv_end}.

We use $\mathcal A_{\textrm{cv},j}=-i\mathcal P(\partial_{k_j}C_{\bs{k}})\Delta_{0,\textrm{cv}}^{-1}$ as a conduction-valence connection with
\be
  \resizebox{0.98\columnwidth}{!}{$
  \displaystyle
C_{\bs{k}}=\begin{pmatrix}
  -k_+/\sqrt6&0&k_-/\sqrt2&0&0&-k_-/\sqrt3\\
  0&-k_+/\sqrt6&0&k_-/\sqrt2&k_+/\sqrt3&0
\end{pmatrix},
  $}
\n
\e
where $k_\pm=k_x\pm ik_y$ and $\Delta_{0,\textrm{cv}}$ is the diagonal matrix of band-edge energy differences between the $\Gamma_{6c}$ conduction states and the $\Gamma_{8v}\oplus\Gamma_{7v}$ valence states. These equations provide all Berry-connection blocks entering the intrinsic source in Eq.~\eqref{eq:coherence_kinetic}. 

In the case of zero spin splitting $\Delta_\textrm{so}=0$, the intrinsic source of the helicity-odd interband coherence reads
\be
\mathrm{Tr}\left[\mathcal{A}_{\textrm{cv},\xi}G_0\mathcal{A}_{\textrm{vv},x}G_0\mathcal{A}_{\textrm{cv},\xi}^{\dagger}\right]=\xi k_y \frac{4\mathcal{P}^4}{3E_\textrm{g}^3} G_0^2,
\e
which makes the Hall character of the intrinsic source explicit.

The source is odd in $k_y$, so its zeroth angular harmonic vanishes in a spatially uniform system, whereas its first transverse moment is finite. The Berry-connection term therefore generates a transverse interband-coherence flux but not a homogeneous helicity density. A boundary is required to convert this odd angular harmonic into the isotropic moment $\mathcal L(y)$ that enters the optical susceptibility.

Electron--hole asymmetry enters at the subsequent propagation stage. The quantity $u_{\rm F}=k_{\rm F}/\mu$ is the relative conduction--valence velocity that transports the optically generated coherence towards the boundary. Thus, equal conduction and valence masses suppress the boundary accumulation even though the odd-in-$k_y$ intrinsic source itself does not vanish.

\textit{Kerr response for finite $\Delta_{\rm so}$.}
For finite valence-band spin-orbit splitting, $\mathcal V_{\textrm{vv},y}$ is no longer proportional to the identity. In addition to the band derivative, it contains off-diagonal terms generated by $[\mathcal A_{\textrm{vv},y},\epsilon_\textrm{v}]$, while the conduction-band connection in Eq.~\eqref{eq:Acc_end} also contributes to the intrinsic source. 

The resulting kinetic equation can still be solved with, for example, specular boundary conditions, but the spatial dependence involves matrix exponentials in the valence subspace. Specifically, the answer would include matrix exponentials of the form $\exp \left[iy (\mathcal{V}_yG_0)^{-1}\right]$, where $\mathcal{V}_y$ is a matrix with non-diagonal blocks originating from $[\mathcal{A}_{\textrm{vv},y},\epsilon_\textrm{v}]$ and constant diagonal entires, which are the relative velocities of the conduction band and the corresponding $\Gamma_{8v}$
 or $\Gamma_{7v}$ valence states.
 
To display the spectral consequence without introducing these lengthy expressions, we evaluate the wide-slab response directly at $y=0$,
\begin{align}
 & \delta\chi^{\rm IC}(0,\omega)=
  -\frac{2e^2\nu_\textrm{F}\mathcal P^2}{3\omega E_\textrm{g}}
  \frac{3eE_{\rm dc}\lambda_\textrm{F}}{2\pi}
  \left(\frac{\mathcal Pk_\textrm{F}}{E_\textrm{g}}\right)^2\nonumber\\
  &\times\left[
  \frac{3\tilde\delta^2+2}{9\tilde\delta^2}
  \frac{1}{\delta_\textrm{F}+i\gamma_\phi}
  +2\frac{\tilde\delta^2+1}{9\tilde\delta^3}
  \frac{1}{\delta_\textrm{F}+\Delta_{\rm so}+i\gamma_\phi}
  \right],
  \label{eq:IC_SO_susceptibility}
\end{align}
where $\tilde\delta=1+\Delta_{\rm so}/E_\textrm{g}$. 

Equation~(\ref{eq:IC_SO_susceptibility}) contains two optical poles associated with transitions into the $\Gamma_{8v}$ and $\Gamma_{7v}$ valence sectors. When $\Delta_{\rm so}\gg\gamma_\phi$, the two contributions are spectrally resolved: the principal resonance occurs near $\delta_{\rm F}=0$, while the split-off contribution becomes resonant near $\delta_{\rm F}=-\Delta_{\rm so}$. The reduction to $5/9$ near the principal resonance should therefore be interpreted as a redistribution of optical weight between the two valence sectors rather than as a requirement of spin--orbit coupling for the effect.

At a finite distance from the boundary, the two transition energies also produce different phase-accumulation scales. Their superposition may lead to a beating structure in the spatial dependence when both transitions lie within the experimentally accessible spectral window. When $\Delta_{\rm so}\lesssim\gamma_\phi$, the two resonances overlap and the response smoothly crosses over to the case of $\Delta_{\rm so}=0$.

%
%
\begin{figure}[!h]
  \centering
  \includegraphics[width=1.0\columnwidth]{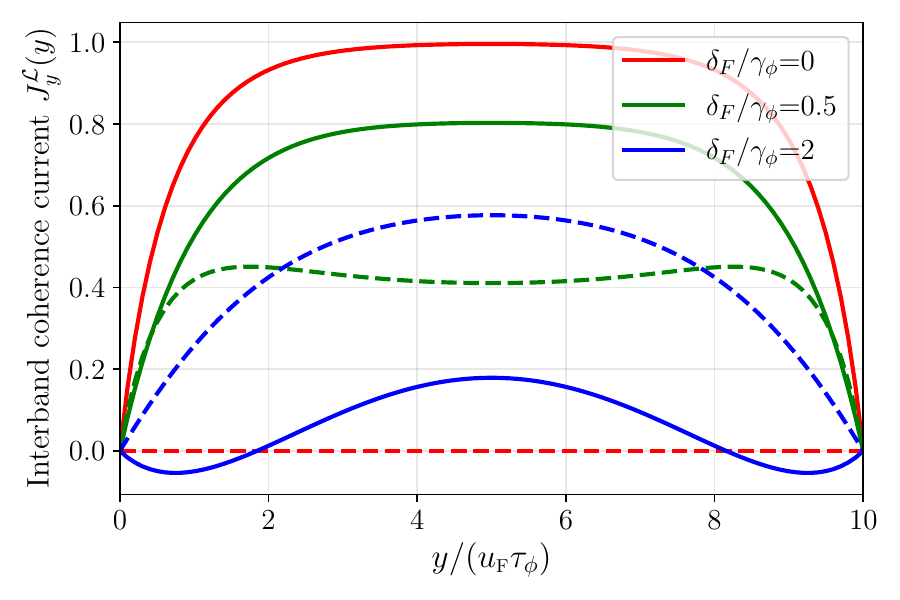}
  \caption{\textbf{Boundary suppression and phase rotation of the interband-coherence flux.}
Real and imaginary parts of $J_y^{\mathcal L}$ are shown by solid and dashed lines, respectively, for three values of $\delta_{\rm F}/\gamma_\phi$. The coordinate is measured in units of the relative conduction--valence dephasing length $u_{\rm F}\tau_\phi$ taking the slab width as $W=10\,u_{\rm F}\tau_\phi$. The coherence flux is normalized to its resonant bulk value. Diffuse boundary conditions enforce $J_y^{\mathcal L}=0$ at both walls. At resonance the current recovers monotonically toward its bulk value, whereas finite detuning produces a spatial phase rotation and sign changes. The flux is always even with respect to the slab center; its gradient generates the antisymmetric coherence density responsible for the Kerr signal.}
  \label{fig:current_profile}
\end{figure}

\textit{Interband-coherence current profile.}
The Kerr signal in Fig.~\ref{fig:spatial} is determined by the isotropic interband coherence density $\mathcal L(y)$. The corresponding transverse coherence flux follows from the same momentum-resolved solution, which is found by imposing diffusive boundary conditions for Eq.~\eqref{eq:L_kinetic}. The general solution for $\mathcal{L}_{\pm, \bf{k}}$ corresponding to positive and negative projections of $k_y$ is given by:
\beml
\begin{align}
k_y>0:&\; \mathcal{L}_{+}(y) = k_y \Gamma + (A_0-k_y\Gamma) e^{-y/\ell_{\bs{k}}},\\
k_y<0:&\; \mathcal{L}_{-}(y) = k_y \Gamma + (A_W-k_y\Gamma) e^{-(W-y)/\ell_{\bs{k}}},
\end{align}
\eml
where $\ell_{\bs{k}} = |k_y|/i\mu G_0$ is trajectory dependent dephasing length, and $\Gamma=2(\mathcal{P}^4/E_\textrm{g}^3)(1-f_\textrm{c})G_0^2$. The constants $A_0$ and $A_W$ are defined as the outgoing boundary amplitudes at $y=0$ and $y=W$, respectively. They are not independent parameters but are determined self-consistently.  Specifically, for the flux-weighted Fuchs conditions of Eq.~(\ref{eq:fuchs_bc}), we find
\begin{align}
&A_{0,W}=\mp 2k \frac{1/3-\En_4\left(W/\ell_{\phi}\right)}{1+2 \En_3\left(W/\ell_{\phi}\right)}\Gamma,
\end{align}
where $k=|\bs{k}|$ is the absolute value of the wave vector. 

For the thick slabs, $W\gg u_\textrm{F}\tau_\phi$, the spatial profile of the coherence flux is determined by the following function
\begin{equation}
  J_y^{\mathcal L}(y)\propto
  2/3-F_\textrm{J}\!\left(y/\ell_\phi\right)
  -F_\textrm{J}\!\left((W-y)/\ell_\phi\right),
  \label{eq:current_profile}
\end{equation}
where $F_\textrm{J}(x)=\En_4(x)+(2/3) \En_3(x)$ and $\En_n$ are defined in Eq.~\eqref{eq:exp_integral}. Equation~\eqref{eq:current_profile} satisfies zero normal coherence flux at both interfaces and approaches the homogeneous bulk flux in the interior of a wide slab.

Figure~\ref{fig:current_profile} shows the real and imaginary parts of the coherence flux $J_y^\mathcal{L}$ from Eq.~\eqref{eq:current_profile}. At resonance, the flux rises monotonically from zero at either wall to its bulk value. Away from the resonance, the coherent phase accumulation reduces the interior coherence flux and can produce local sign changes close to the interfaces.

\end{document}